# Stochastic Analysis of the Diffusion Least Mean Square and Normalized Least Mean Square Algorithms for Cyclostationary White Gaussian and Non-Gaussian Inputs

Eweda Eweda[1],  Neil J. Bershad[2],  Jose C. M. Bermudez[3]

[1] Department of Electrical Engineering, Future University, Cairo, Egypt (e-mail: eweda@ieee.org).
[2] Department of Electrical Engineering and Computer Science, University of California, Irvine, CA 92660 USA (e-mail: bershad@ece.uci.edu).
[3] Department of Electrical Engineering, Federal University of Santa Catarina, Florianópolis, SC, Brazil (e-mail: j.bermudez@ieee.org).

**Abstract**

The diffusion least mean square (DLMS) and the diffusion normalized least mean square (DNLMS) algorithms are analyzed for a network having a fusion center. This structure reduces the dimensionality of the resulting stochastic models while preserving important diffusion properties. The analysis is done in a system identification framework for cyclostationary white nodal inputs. The system parameters vary according to a random walk model. The cyclostationarity is modeled by periodic time variations of the nodal input powers. The analysis holds for all types of nodal input distributions and nodal input power variations. The derived models consist of simple scalar recursions. These recursions facilitate the understanding of the network mean and mean-square dependence upon the 1) nodal weighting coefficients, 2) nodal input kurtosis and cyclostationarities, 3) nodal noise powers and 4) the unknown system mean-square parameter increments. Optimization of the node weighting coefficients is studied. Also investigated is the stability dependence of the two algorithms upon the nodal input kurtosis and weighting coefficients. Significant differences are found between the behaviors of the DLMS and DNLMS algorithms for non-Gaussian nodal inputs. Simulations provide strong support for the theory.

**KEYWORDS:**

Adaptive Filters, Least Mean Square Algorithm, Normalized Least Mean Square Algorithm, Stochastic Diffusion Algorithms

# 1 INTRODUCTION

Diffusion algorithms [1-12] and cyclostationary inputs [13-23] are two separate important subjects in the theory of adaptive filtering. The purpose of this paper is to combine the two subjects and to analyze the stochastic behavior of diffusion algorithms for cyclostationary inputs. There have been a large number of papers published in the last 15-20 years on the subject of adaptive diffusion algorithms, i.e. algorithms whose inputs consist of multiple observations of similar phenomena. Professor Ali Sayed and his co-researchers have published a large variety of structural and stochastic analysis papers on the subject [1-12].

Examples of applications involving cyclostationary signals are mentioned in [22, p. 4753]. Existing studies of diffusion algorithms for cyclostationary inputs [24-26] involved a vector structure which gives rise to complex recursions for the adaptive weight error vector and the resultant mean square deviation (MSD). They use a Kronecker product formulation introduced in [6], resulting in block diagonal matrices of the size $NM \times NM$ where $N$ is the adaptive filter length and $M$ is the number of nodes. The $NM \times NM$ diagonal matrices are very cumbersome and provide neither any insight to the design of the algorithm nor much physical insight into what is happening as the parameters of the DLMS algorithm are varied. Ref. [24] attempted to analyze a diffusion signed LMS algorithm. However, several unjustified and questionable assumptions involving replacing random quantities by their means have been used to overcome the complexities of the analysis [24-eqs. (49) and (50)]. Ref. [25] analyzed the diffusion LMS (DLMS) algorithm for cyclostationary inputs using energy principles [27, 28] and a Kronecker product formulation. There are two problems with this approach: 1) the energy principle [27, 28] is only applicable if the filter converges, which is not the case for cyclostationary inputs; 2) the use of the Kronecker product formulation with the disadvantages listed above. Finally, [26] studied the case of the DLMS algorithm for cyclostationary white non-Gaussian inputs. This is an extension of the work in [22] to the DLMS algorithm, which also uses Kronecker product formulation with the disadvantages listed above. In summary, the theory in [24-26] makes analysis and design very difficult and reduces the advantages provided by generality. Most of the time, practical designs are based on the study of simplified structures.

The present paper considers one structural simplification. In this structure, the network consists of a set of nodes and a fusion center (FC). This simplification significantly reduces the dimensionality of the resulting models while preserving important diffusion properties. Therefore, the resulting models facilitate the understanding of the effect of diffusion on the network mean and mean-square behaviors for cyclostationary inputs of the nodes. The resulting models also facilitate the understanding of dependence upon the parameters of the network, and provide insights into the design of the general DLMS algorithm. The considered simplified structure can also exist in practice. Namely, it can define a tactical situation where a command center can communicate with the nodes but the nodes are unable or unwilling to communicate with each other. Since the network has an FC, the corresponding diffusion LMS and NLMS algorithms will be termed as the FC-DLMS and FC-DNLMS algorithms respectively.

The present paper provides mean and mean-square analysis of the FC-DLMS and FC-DNLMS algorithms in a system identification framework for cyclostationary white nodal inputs. The system parameters vary according to a random walk model. The cyclostationarity is modeled by periodic time variations of the nodal input powers. The analysis holds for 1) all types of nodal input distributions and 2) all rates of variations of the nodal input powers. These two points are important for adaptive networks, since different nodes generally have different types of input distribution and different rates of variation of the input powers. Simple scalar recursions are derived for the mean and MSD of the FC. The obtained model enables easy understanding of the dependence of the FC mean and mean-square behaviors on the weighting coefficients of the nodes, kurtosis and cyclostationarities of the nodal inputs, nodal noise powers and mean square increment of the unknown system. Optimization of the node weighting coefficients is studied. Dependence of the stability of the algorithms on the nodal input kurtosis and node weighting coefficients is also investigated. Significant differences are found between the behaviors of the DLMS and DNLMS algorithms for non-Gaussian nodal inputs. Monte Carlo simulations are in perfect agreement with the theory. This high accuracy of the derived models follows from the fact that they are almost free of approximations, as will be seen in Section 3.

The paper is organized as follows. Section 2 defines the problem and the used statistical assumptions. Section 3 derives simple recursions for the mean and MSD of the FC for the DLMS algorithm. Section 4 studies the dependence of the mean and mean square behaviors of the DLMS algorithm on the node input kurtosis and node weighting coefficients. Section 5 is devoted to the analysis of the DNLMS algorithm. Section 6 compares the theory to Monte Carlo simulations. Conclusions are presented in Section 7.

## 2 FORMULATION OF THE PROBLEM

### 2.1 The FC-DLMS Algorithm

The FC-DLMS algorithm is defined as an adaptive filtering algorithm for a network consisting of $M$ nodes and a fusion center (FC). The nodes cooperate in estimating the weight vector $H(n)$ of an unknown linear time-varying system; $n$ is the discrete time index. Each node has access to the input and output of the system. The input and output signals corresponding to the $j$-th node are respectively denoted by $x_j(n)$ and $d_j(n)$, where

$$d_j(n) = X_j^T(n)H(n) + n_j(n) \tag{1}$$

where $X_j(n) = [x_j(n), x_j(n-1), \ldots, x_j(n-N+1)]^T$, the superscript $T$ means transpose and $n_j(n)$ is the measurement noise. Each node has a local LMS-type adaptive filter that sends a weighted local estimate of $H(n)$ to the FC. There are two strategies for adapting the network [1]: 1) The Combine Then Adapt (CTA) strategy and 2) The Adapt Then Combine (ATC) strategy.

The CTA FC-DLMS algorithm is described by the two following equations

$$\theta(n) = \sum_{k=1}^{M} c_k W_k(n) \tag{2}$$

$$W_j(n+1) = \theta(n) + \mu_j[d_j(n) - X_j^T(n)\theta(n)]X_j(n). \tag{3}$$

This algorithm is depicted in Figure 1. Two steps are performed at time *n*. The first step is the combine step given by eq. (2). The FC receives the nodal local weight estimates $W_k(n)$ and combines them to produce the common estimate $\theta(n)$ and sends it to all nodes. The adapt step is the second step given by eq. (3). The local LMS adaptive filters adapt using $\theta(n)$ to produce their new local weight estimates $W_j(n+1)$. $c_k > 0$ is the weighting coefficient for node *k* in eq. (2). These coefficients satisfy

$$\sum_{k=1}^{M} c_k = 1. \tag{4}$$

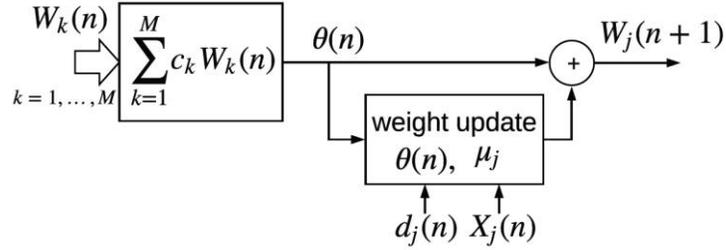

Figure 1: CTA FC-DLMS

The ATC FC-DLMS algorithm is described by the two following equations

$$\theta_j(n+1) = W(n) + \mu_j[d_j(n) - X_j^T(n)W(n)]X_j(n) \tag{5}$$

$$W(n+1) = \sum_{k=1}^{M} c_k \theta_k(n+1). \tag{6}$$

This algorithm is depicted by Figure 2. Two steps are performed at time *n*. The first step is the adapt step. The local LMS algorithms adapt using the common weight estimate *W(n)* as given in eq. (5) and send their obtained intermediate estimates $\theta_j(n+1)$ to the FC. The combine step is the second step. The FC combines the intermediate estimates to produce the new common weight estimate *W(n+1)* and sends it to all nodes as given in eq. (6).

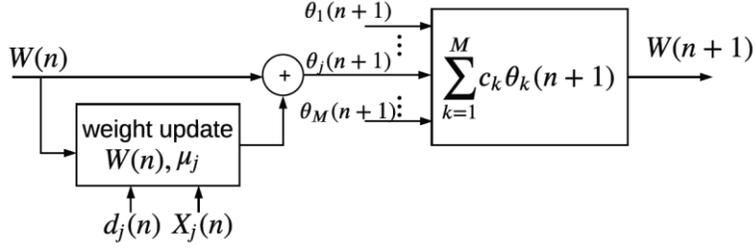

Figure 2: ATC FC-DLMS

## 2.2 Cyclostationary Nodal Input Signal Models

The *j*-th node input signal is modelled by

$$x_j(n) = \sigma_{xj}(n) s_j(n) \tag{7}$$

where $\sigma_{xj}^2(n)$ is a deterministic periodic sequence with period $T_j$ and $s_j(n)$ is a zero-mean unity variance i.i.d. random sequence. Equation (7) is a typical form of cyclostationary signal used in many practical applications, such as communications through flat-fading channels, radar, sonar, frequency modulation, fractionally-sampled communications, channel estimation, etc. [29-31]. Thus, $X_j(n)$ is a zero-mean white vector with time-varying variance such that

$$R_{Xj}(n) \equiv E[X_j(n) X_j^T(n)] = Diag[\sigma_{xj}^2(n), \sigma_{xj}^2(n-1), \ldots, \sigma_{xj}^2(n-N+1)]. \tag{8}$$

Hence, $X_j(n)$ is a discrete time wide sense cyclostationary process.

It is well known [32] that white models are often not accurate models of real world signals. However, analytical predictions based on these white models are often very useful in predicting the behavior for the real world signals. Although the above model is not general, (8) defines a non-trivial model. The input displays a simple type of cyclostationarity which can be used to model more complex time-varying input statistical properties. More importantly, the behaviors of the FC-DLMS and FC-DNLMS algorithms can be accurately analyzed for this input signal as will be shown subsequently. Two simple models for $\sigma_{xj}^2(n)$ are considered here: a sinusoidal power time variation

$$\sigma_{xj}^2(n) = \beta_j (1 + \sin(\omega_{oj} n)) \tag{9}$$

and a pulsed power time variation

$$\sigma_{xj}^2(n) = \begin{cases} P_{1j} & \text{for } iT_j < n \leq iT_j + \alpha_j T_j \\ P_{2j} & \text{for } iT_j + \alpha_j T_j < n \leq (i+1)T_j \end{cases}, \quad 0 < \alpha_j < 1, \, i = 1, 2, \ldots. \tag{10}$$

The theory presented here can be extended to other cyclostationary power variations in a similar manner.

The sinusoidal variation model can be used to study the algorithm behavior for different speeds of input power variation with bounded maximum power. The pulsed model can be used to study the algorithm behavior for pulsed inputs such as those occurring in digital data transmission and pulsed radar systems [29, 33]. These independent power variations provide new insights into the behavior of the algorithms with more general inputs.

The time variations of the input power can be classified as slow, moderate or fast depending on how the variation period compares to the filter memory, i.e. to the number of taps $N$. Hence, the variations are slow for (9) if $\omega_{oj} N \ll 2\pi$ and for (10) if $N \ll T_j$. The variations are fast for (9) if $\omega_{oj} N \gg 2\pi$ and for (10) if $N \gg T_j$. The variations are moderate for (9) if $\omega_{oj} N \approx 2\pi$ and for (10) if $N \approx T_j$. These conditions are different than those relating the input power time variations to the time constants (memory) of the algorithms.

It is also assumed that different nodal inputs are mutually independent; i.e. the sequences $x_j(n)$ and $x_k(n)$ are mutually independent for $j \neq k$. The noise sequence $n_j(n)$ is assumed a white sequence independent of all other sequences in the network.

The standard random walk model is used for the unknown system

$$H(n+1) = H(n) + Q(n) \tag{11}$$

where the system parameter increment vector $Q(n)$ is a white vector, i.e. $E[Q(n)Q^T(n)] = \sigma_q^2 I$, where $I$ is the identity matrix. The vector sequence $Q(n)$ is assumed independent of all other sequences in the network. The random-walk model (11) is not realistic. However, this model is widely adopted in adaptive filtering literature since it allows a feasible tracking analysis. This model is an approximation to the first order Markov model [27,28].

Finally, we employ the Independence Theory (IT) of adaptive filtering, which assumes that the weights at time *n* are statistically independent of the input vector at time *n* [27]. The use of this assumption considerably simplifies the stochastic analysis of the adaptive filter. The IT assumption has been shown to lead to very accurate models in a wide variety of adaptive filter applications.

## 3 STOCHASTIC ANALYSIS OF THE FUSION CENTER

### 3.1 Combine-Then-Adapt Strategy (CTA)

Subtracting $H(n)$ from the two sides of (3), using (1) and (11), it follows that

$$V_j(n+1) = P(n) - \mu_j X_j(n)[X_j^T(n)P(n) - n_j(n)] - Q(n), \quad j = 1, 2, \ldots, M \tag{12}$$

where $V_j(n) \equiv W_j(n) - H(n)$ and

$$P(n) \equiv \sum_{j=1}^{M} c_j V_j(n) = \sum_{j=1}^{M} c_j W_j - H(n). \tag{13}$$

Multiplying the two sides of (12) by $c_j$, summing over *j* and using (4), we obtain

$$P(n+1) = P(n) - \sum_{j=1}^{M} c_j \mu_j X_j(n) X_j^T(n) P(n) + \sum_{j=1}^{M} c_j \mu_j n_j(n) X_j(n) - Q(n). \tag{14}$$

### 3.1.1 Mean behavior of (14)

The IT and (13) imply that $P(n)$ and $X_j(n)$ are mutually independent. Hence, averaging both sides of (14) yields a recursion for $E\{P(n)\}$,

$$E\{P(n+1)\} = \left[I - \sum_{j=1}^{M} c_j \mu_j R_{Xj}(n)\right] E\{P(n)\}. \tag{15}$$

Further progress with (15) requires specifying $R_{Xj}(n)$ for all *j*. This is considered in Section 4.

### 3.1.2 MSD behavior of *P(n)*

In this section, we use the techniques in [22] to derive an MSD model for the FC-DLMS algorithm. The derived model provides many interesting conclusions about the effect of diffusion on the network mean-square behavior for cyclostationary white Gaussian and non-Gaussian inputs of the nodes.

Multiplying (14) from the right by its transpose yields

$P(n+1)P^T(n+1) =$

$P(n)P^T(n) - \sum_{j=1}^{M} c_j\mu_j X_j(n) X_j^T(n)P(n)P^T(n) - \sum_{j=1}^{M} c_j\mu_j P(n)P^T(n)X_j(n)X_j^T(n) +$

$\sum_{j=1}^{M}\sum_{k=1}^{M} c_j\mu_j c_k\mu_k X_j(n)X_j^T(n)P(n)P^T(n)X_k(n)X_k^T(n) +$

$\sum_{j=1}^{M}\sum_{k=1}^{M} c_j\mu_j n_j(n) c_k\mu_k n_k(n) X_j(n) X_k^T(n) + Q(n) Q^T(n)$

+ *four statistically independent cross terms which will average to zero*. (16)

We need to average both sides of (16). Now,

$E\{\sum_{j=1}^{M}\sum_{k=1}^{M} c_j\mu_j c_k\mu_k X_j(n)X_j^T(n)P(n)P^T(n)X_k(n)X_k^T(n)\} =$

$E\{\sum_{j=1}^{M} c_j^2\mu_j^2 X_j(n)X_j^T(n)P(n)P^T(n)X_j(n)X_j^T(n)\} +$

$E\{\sum_{j=1}^{M}\sum_{k=1,k\neq j}^{M} c_j\mu_j c_k\mu_k X_j(n)X_j^T(n)P(n)P^T(n)X_k(n)X_k^T(n)\}.$ (17)

Denote

$$K(n) = E\{P(n)P^T(n)\} \tag{18}$$

and denote the *ii*-th entry of $K(n)$ by $K_{ii}(n)$. We shall derive a recurrence for $K_{ii}(n)$. From (16), we have

$$K_{ii}(n+1) = K_{ii}(n) - 2K_{ii}(n)\sum_{j=1}^{M} c_j\mu_j\sigma_{xj}^2(n-i+1) + T_1(n) + T_2(n) + \sum_{j=1}^{M} c_j^2\mu_j^2\sigma_{nj}^2\sigma_{xj}^2(n-i+1) + \sigma_q^2 \tag{19}$$

where $T_1(n)$ and $T_2(n)$ are the *ii*-th entries of the first and second terms on the RHS of (17) respectively.

Hence,

$$T_1(n) = \sum_{j=1}^{M} c_j^2\mu_j^2 E\{x_j^2(n-i+1)[X_j^T(n)P(n)]^2\}. \tag{20}$$

The IT and the mutual independence of the entries of $X_j(n)$ imply that

$$E\{x_j^2(n-i+1)[X_j^T(n)P(n)]^2\} = E\{x_j^4(n-i+1)\}K_{ii}(n) + \sum_{r=1,r\neq i}^{N}\sigma_{xj}^2(n-i+1)\sigma_{xj}^2(n-r+1)K_{rr}(n). \tag{21}$$

Denote the input kurtosis at node *j* by $\psi_j$. Then, (21) implies that

$$E\{x_j^2(n-i+1)[X_j^T(n)P(n)]^2\} = \psi_j\sigma_{xj}^4(n-i+1)K_{ii}(n) + \sum_{r=1,r\neq i}^{N}\sigma_{xj}^2(n-i+1)\sigma_{xj}^2(n-r+1)K_{rr}(n). \tag{22}$$

Inserting (22) in (20), we obtain

$$T_1(n) = K_{ii}(n)\sum_{j=1}^{M} c_j^2 \mu_j^2 \psi_j \sigma_{xj}^4(n-i+1) + \sum_{j=1}^{M} c_j^2 \mu_j^2 \sum_{r=1, r\neq i}^{N} \sigma_{xj}^2(n-i+1)\sigma_{xj}^2(n-r+1)K_{rr}(n). \tag{23}$$

Now, we consider $T_2(n)$, the $ii$-th entry of the second term on the RHS of (17). The mutual independence of $X_j(n)$ and $X_k(n)$, the mutual independence of the entries of $X_j(n)$, and IT imply that

$$T_2(n) = K_{ii}(n)\sum_{j=1}^{M}\sum_{k\neq j}^{M} c_j \mu_j c_k \mu_k \sigma_{xj}^2(n-i+1)\sigma_{xk}^2(n-i+1). \tag{24}$$

Inserting (23) and (24) in (19), we obtain

$$K_{ii}(n+1) = K_{ii}(n) - \alpha_i(n)K_{ii}(n) + \sum_{r=1, r\neq i}^{N} \beta_{ir}(n)K_{rr}(n) + \gamma_i(n) \tag{25}$$

where

$$\alpha_i(n) = 2\sum_{j=1}^{M} c_j \mu_j \sigma_{xj}^2(n-i+1) - \sum_{j=1}^{M} c_j^2 \mu_j^2 \psi_j \sigma_{xj}^4(n-i+1)$$

$$-\sum_{j=1}^{M}\sum_{k\neq j}^{M} c_j \mu_j c_k \mu_k \sigma_{xj}^2(n-i+1)\sigma_{xk}^2(n-i+1) \tag{26}$$

$$\beta_{ir}(n) = \sum_{j=1}^{M} c_j^2 \mu_j^2 \sigma_{xj}^2(n-i+1)\sigma_{xj}^2(n-r+1) \tag{27}$$

$$\gamma_i(n) = \sum_{j=1}^{M} c_j^2 \mu_j^2 \sigma_{nj}^2 \sigma_{xj}^2(n-i+1) + \sigma_q^2. \tag{28}$$

The MSD of the FC is given by

$$MSD(n) = \sum_{i=1}^{N} K_{ii}(n). \tag{29}$$

The MSD model given by (25)-(29) consists of $N$ scalar recursions. Hence, it is easily computable. Monte Carlo simulations (see Section 4) are in excellent agreement with this model. The model is valid for general distributions of the node inputs, including the Gaussian distribution. Also, the model is valid for all rates of variation of the input powers of the nodes. Eqs. (25)-(29) show that the MSD behavior of the FC-DLMS algorithm depends on the node input distribution through only the kurtosis. This implies that different input distributions with the same kurtosis yield the same MSD for the FC-DLMS algorithm. Equation (26) implies that the MSD dependence on the type of input distribution is negligible for small $\mu$ and that the dependence increases as $\mu$ increases, since the kurtosis appears only in the $\mu^2$ term in the MSD recurrence. This is further detailed in part B of Section 4 below.

### 3.2 Adapt-Then-Combine Strategy (ATC)

The ATC recursions are given by (5) and (6). Inserting (1) in (5) yields

$$\theta_k(n+1) = W(n) + \mu_k X_k(n)\big[X_k^T(n)H(n) + n_k(n) - X_k^T(n)W(n)\big]. \tag{30}$$

Weighting (30) by $c_k$, summing on $k$, using (4) and (6), it follows that

$$W(n+1) = W(n) + \sum_{k=1}^{N} c_k \mu_k X_k(n)\big[X_k^T(n)H(n) + n_k(n) - X_k^T(n)W(n)\big]. \tag{31}$$

Subtracting $H(n)$ from both sides of (31), using (11), and defining $P_{ATC}(n) = W(n) - H(n)$ yields

$$P_{ATC}(n+1) = P_{ATC}(n) - \sum_{k=1}^{M} c_k \mu_k X_k(n) X_k^T(n) P_{ATC}(n) + \sum_{k=1}^{N} c_k \mu_k X_k(n) n_k(n) - Q(n). \tag{32}$$

The recursion (32) is identical with the recursion for $P(n)$ in (14). Consequently, the fusion center output deviation of the ATC, $P_{ATC}(n+1)$, is the same as the fusion center output deviation of CTA, $P(n+1)$. Hence, the theory and MC simulations are the same for both CTA and ATC strategies.

The models derived in this section are valid for any combination of node input frequencies, input powers, input distributions and step sizes. In the following we study the performance dependence on the input statistics, and derive relevant properties of the algorithms in the especially important cases, such as uniform weighting (same $c_j$ for all $j$), or same adaptation speed for all nodes. These cases are very simple to design due to the reduced number of parameters. Therefore, understanding their performance is of great practical interest.

## 4 PERFORMANCE DEPENDENCE ON THE KURTOSIS AND NODE WEIGHTS

In this section, equations are derived for the analysis of the dependence of the MSD behavior on the kurtosis and node weights. For mathematical tractability, this analysis is done for the case of slow input power variations, which is important in practice.

## 4.1 Behavior for Slowly-Varying Node Input Powers

For slowly varying input power, the filter length will be small in comparison with the period of variation of the input power. Consequently,

$$\sigma_{xj}^2(n) \approx \sigma_{xj}^2(n-1) \approx ... \approx \sigma_{xj}^2(n-N+1). \tag{33}$$

From (8) and (33), we have

$$R_{Xj}(n) \approx \sigma_{xj}^2(n)I. \tag{34}$$

### 4.1.1 Mean weight behavior

Inserting (34) in (15), we obtain

$$E\{P(n+1)\} = \left[1 - \sum_{j=1}^{M} c_j \mu_j \sigma_{xj}^2(n)\right] E\{P(n)\}. \tag{35}$$

From (35), the mean weight behavior does not depend on the node input distribution type. The mean weight behavior depends on the node weights. For wide-sense stationary node inputs, i.e. $\sigma_{xj}^2(n) = \sigma_{xj}^2$, the CF-DLMS algorithm is mean stable if and only if

$$0 < \sum_{j=1}^{M} c_j \mu_j \sigma_{xj}^2 < 2.$$

This is further discussed in Section 4.2 below.

### 4.1.2 MSD behavior

From (26)-(28) and (33), $\alpha_i(n)$, $\beta_{ir}(n)$ and $\gamma_i(n)$ do not depend on $i$ or $r$. Denote them by $\alpha(n)$, $\beta(n)$ and $\gamma(n)$ respectively. Hence, (25) becomes

$$K_{ii}(n+1) = K_{ii}(n) - [\alpha(n) + \beta(n)]K_{ii}(n) + \beta(n)MSD(n) + \gamma(n) \tag{36}$$

where $MSD(n)$ is given by (29) and

$$\alpha(n) = 2\sum_{j=1}^{M} c_j \mu_j \sigma_{xj}^2(n) - \sum_{j=1}^{M} c_j^2 \mu_j^2 \psi_j \sigma_{xj}^4(n) - \sum_{j=1}^{M}\sum_{k \neq j}^{M} c_j \mu_j c_k \mu_k \sigma_{xj}^2(n) \sigma_{xk}^2(n) \tag{37}$$

$$\beta(n) = \sum_{j=1}^{M} c_j^2 \mu_j^2 \sigma_{xj}^4(n) \tag{38}$$

$$\gamma(n) = \sum_{j=1}^{M} c_j^2 \mu_j^2 \sigma_{nj}^2 \sigma_{xj}^2(n) + \sigma_q^2. \tag{39}$$

Summing (36) over $i$, we obtain

$$MSD(n+1) = [1 - \alpha(n) + (N-1)\beta(n)]MSD(n) + N\gamma(n). \tag{40}$$

From (37) and (38), we have

$$-\alpha(n) + (N-1)\beta(n) = -2\sum_{j=1}^{M} c_j \mu_j \sigma_{xj}^2(n) + \sum_{j=1}^{M} c_j^2 \mu_j^2 \sigma_{xj}^4(n)(\psi_j + N - 2) + \left[\sum_{j=1}^{M} c_j \mu_j \sigma_{xj}^2(n)\right]^2. \tag{41}$$

From (39)-(41), the MSD recurrence of the FC-DLMS algorithm is given by

$$MSD(n+1) = \left\{1 - 2\sum_{j=1}^{M} c_j \mu_j \sigma_{xj}^2(n) + \sum_{j=1}^{M} c_j^2 \mu_j^2 \sigma_{xj}^4(n)(\psi_j + N - 2) + \left[\sum_{j=1}^{M} c_j \mu_j \sigma_{xj}^2(n)\right]^2\right\} MSD(n)$$

$$+ N\sum_{j=1}^{M} c_j^2 \mu_j^2 \sigma_{nj}^2 \sigma_{xj}^2(n) + N\sigma_q^2. \tag{42}$$

Equation (42) implies that slow input power variations will cause MSD ripples in the transient phase. The ripple repetition period is equal to the least common multiple of the power repetition periods at all nodes. Equation (42) also implies that the MSD dependence on the kurtosis decreases as $N$ increases.

Now, let us investigate the steady-state behavior of the algorithm. For time-varying input power, the term steady-state is used to describe the algorithm behavior after the transient has disappeared. It does not imply a time invariant quantity. We assume that the fluctuations in $MSD(n)$ are small enough in the steady state so that we can use the approximation $MSD(n+1) \approx MSD(n)$. Hence, (42) implies that

$$MSD_{steady-state}(n) = \left\{2\sum_{j=1}^{M} c_j \mu_j \sigma_{xj}^2(n) - \sum_{j=1}^{M} c_j^2 \mu_j^2 \sigma_{xj}^4(n)(\psi_j + N - 2) - \left[\sum_{j=1}^{M} c_j \mu_j \sigma_{xj}^2(n)\right]^2\right\}^{-1}$$

$$\times \left\{N\sum_{j=1}^{M} c_j^2 \mu_j^2 \sigma_{nj}^2 \sigma_{xj}^2(n) + N\sigma_q^2\right\}. \tag{43}$$

Eq. (43) implies that slow input power variations generally cause ripples in the steady-state MSD. The ripple repetition period is equal to the least common multiple of the power repetition periods at all nodes.

## 4.2 Case of Wide Sense Stationary Node Inputs

To further investigate the effects of kurtosis and node weighting on the behavior of the FC-DLMS algorithm, we shall use some simplifying assumptions. We shall assume that $\sigma_{xj}^2(n)$ does not depend on $n$; denote it by $\sigma_{xj}^2$. This assumption implies that the input is wide sense stationary process, which is a special case of cyclostationarity. Thus, FC results, based on this assumption, can be used to better understand the general case studied previously, e.g. [9], using the Kronecker product formulation.

As is usually the case in practice, we shall assume that $\mu_j$ is proportional to $1/\sigma_{xj}^2$; i.e.

$$\mu_j = \frac{\lambda}{\sigma_{xj}^2} \tag{44}$$

where $\lambda$ is a constant being the same for all $j$, so that all node filters adapt at the same speed if all inputs have equal kurtosis. This constant will be termed as the normalized step-size. We shall also assume that all nodes have the same type of input distribution; i.e. $\psi_j$ does not depend on $j$; denote it by $\psi$. This assumption is used to obtain a simple MSD model for better understanding the algorithm behavior.

The optimum weighting coefficients $c_j$ are usually chosen on the basis of minimizing the steady-state MSD and are dependent upon the relative nodal signal to noise ratios (SNR). If one of the nodal SNR's is dominant, then one would expect that its weight would be near one. However, as will be shown below, the optimum weights are uniform (1/M for all nodes) instead.

Using (4), (42) and (44) we obtain

$$MSD(n+1) = \left\{ 1 - 2\lambda + (N + \psi - 2)\lambda^2 \sum_{j=1}^{M} c_j^2 + \lambda^2 \right\} MSD(n) + N\lambda^2 \sum_{j=1}^{M} \frac{c_j^2}{\rho_j} + N\sigma_q^2. \tag{45}$$

where

$$\rho_j \equiv \frac{\sigma_{xj}^2}{\sigma_{nj}^2} \tag{46}$$

is the SNR at node $j$.

Equation (45) implies that the MSD dependence on the kurtosis is increasing in the normalized step-size $\lambda$ and in the sum of the squared node weights. The convergence rate of the algorithm is decreasing in the kurtosis and in the sum of the squared node weights. Thus, the fastest convergence is attained at the minimum value of $\sum_{j=1}^{M} c_j^2$. From (4) and the Appendix, with $\eta_j = 1$, this minimum is attained when all weights are equal; i.e. uniform weighting. In such a case, $\sum_{j=1}^{M} c_j^2 = \frac{1}{M}$. For this case, the dependence on the kurtosis decreases as $M$ increases.

For small values of the normalized step-size $\lambda$, eq. (45) implies that the dependence on the kurtosis is negligible for designs satisfying (44). The steady-state MSD in this case is given by

$$MSD(\infty) = \frac{1}{2}\left[ N\lambda \sum_{j=1}^{M} \frac{c_j^2}{\rho_j} + N\lambda^{-1}\sigma_q^2 \right]. \tag{47}$$

From (4) and the Appendix, with $\eta_j = \rho_j$, the minimum value of $\sum_{j=1}^{M} \frac{c_j^2}{\rho_j}$ is attained at

$$c_j = \frac{\rho_j}{\rho_1 + \rho_2 + \ldots + \rho_M}. \tag{48}$$

This equation gives the node weights that minimize the steady-state MSD for small values of the normalized step-size $\lambda$. In this case, the minimum steady-state MSD is given by

$$MSD(\infty)_{\min} = \frac{1}{2}\left[ \frac{N\lambda}{\rho_1 + \rho_2 + \ldots + \rho_M} + N\lambda^{-1}\sigma_q^2 \right]. \tag{49}$$

Eq. (49) implies that the steady-state MSD is decreasing as the number of nodes increases and the node signal-to-noise ratios increase. Both behaviors agree with physical intuition.

This section is terminated by a stability analysis of the FC-DLMS algorithm. From (35) and (44), we have

$$E\{P(n+1)\} = [1 - \lambda]E\{P(n)\}. \tag{50}$$

This equation implies that the FC mean weight deviation is stable if and only if the normalized step-size $\lambda$ satisfies the condition

$$0 < \lambda < 2. \tag{51}$$

Eq. (45) implies that the necessary and sufficient condition of the MSD stability is given by

$$0 < \lambda < \frac{2}{1+(N+\psi-2)\sum_{j=1}^{M} c_j^2}. \tag{52}$$

From (52), the step-size stability bound depends on the kurtosis and on the node weights. The maximum step-size stability bound is attained at the minimum value of $\sum_{j=1}^{M} c_j^2$, which takes place when $c_j = 1/M$ for all $j$. In this case, the stability range is given by

$$0 < \lambda < \frac{2M}{M+(N+\psi-2)}. \tag{53}$$

Eq. (53) implies that the step-size stability bound increases as the number of nodes increases. Eq. (53) has a very interesting implication. For a single node, $M=1$ and (53) yields $0 < \lambda < 2/(N+\psi-1)$. Thus, the stability range for a single node is less than for the FC. Hence the convergence rate for the FC can be faster than for a single node without causing instability. In essence, each of the nodes is helping to stabilize the other nodes. The effects of this property on the MSD will be investigated in Section 6 when comparing the theory and Monte Carlo simulations.

## 5 ANALYSIS OF THE FC-DNLMS ALGORITHM

### 5.1 Analysis

In the FC-DNLMS algorithm, the local node adaptive filters are NLMS, rather than LMS, adaptive filters. Hence, from (2) and (3), the CTA FC-DNLMS algorithm is described by the two following equations

$$\theta(n) = \sum_{k=1}^{M} c_k W_k(n) \tag{54-a}$$

$$W_j(n+1) = \theta(n) + \frac{\xi_j}{X_j^T(n)X_j(n)}\left[d_j(n) - X_j^T(n)\theta(n)\right]X_j(n) \tag{54-b}$$

where $\xi_j > 0$ is the step-size of the NLMS algorithm at the $j$-th node.

From (5) and (6), the ATC FC-DNLMS algorithm is described by the two following equations

$$\theta_j(n+1) = W(n) + \frac{\xi_j}{X_j^T(n)X_j(n)}\left[d_j(n) - X_j^T(n)W(n)\right]X_j(n) \tag{55}$$

$$W(n+1) = \sum_{k=1}^{M} c_k \theta_k(n+1). \tag{56}$$

Straightforward extension of Section 3.2 shows that the fusion center output of the ATC FC-DNLMS algorithm is the same as the fusion center output of the CTA FC-DNLMS algorithm. Consequently, the analysis in this section will be presented for only one of the two strategies, namely the CTA.

Using (54) and following the same procedure as the derivation of (14), we obtain

$$P(n+1) = P(n) - \sum_{j=1}^{M} \frac{c_j \xi_j X_j(n) X_j^T P(n)}{X_j^T(n)X_j(n)} + \sum_{j=1}^{M} \frac{c_j \xi_j n_j(n) X_j(n)}{X_j^T(n)X_j(n)} - Q(n) \tag{57}$$

where

$$P(n) \equiv \sum_{j=1}^{M} c_j V_j(n), \quad V_j(n) \equiv W_j(n) - H(n). \tag{58}$$

From [22, eqs. (34), (35)], we have

$$X_j^T(n)X_j(n) \approx N\overline{\sigma_{xj}^2(n)} \text{ for } N \gg \psi_j \tag{59}$$

where

$$\overline{\sigma_{xj}^2(n)} \equiv \frac{1}{N}\sum_{i=1}^{N}\sigma_{xj}^2(n-i+1). \tag{60}$$

Throughout this section, it is assumed that

$$N \gg \psi_j \text{ for all } j. \tag{61}$$

Note that this assumption is not used for the case of the FC-DLMS algorithm. Inserting (60) in (57), we obtain

$$P(n+1) = P(n) - \sum_{j=1}^{M} \frac{c_j \xi_j X_j(n) X_j^T(n) P(n)}{N\overline{\sigma_{xj}^2(n)}} + \sum_{j=1}^{M} \frac{c_j \xi_j n_j(n) X_j(n)}{N\overline{\sigma_{xj}^2(n)}} - Q(n). \tag{62}$$

Comparing the FC-DNLMS recurrence (62) with the FC-DLMS recurrence (14) and using the fact that the sequence $\overline{\sigma_{xj}^2(n)}$ is deterministic, it follows that the mean and mean square behaviors of the FC-DNLMS algorithm can be obtained from those of the FC-DLMS algorithm after using the following substitution

$$\mu_j = \frac{\xi_j}{N\overline{\sigma_{xj}^2(n)}}. \tag{63}$$

**5.2 Behavior for General Rates of Variation of the Node Input Powers**

The mean weight behavior of the FC-DNLMS algorithm is obtained by inserting the substitution (63) in (15), which yields

$$E\{P(n+1)\} = \left[I - \sum_{j=1}^{M} \frac{c_j \xi_j R_{Xj}(n)}{N\overline{\sigma_{xj}^2(n)}}\right] E\{P(n)\}. \tag{64}$$

The MSD recurrence is obtained from (25) after inserting the substitution (63) in (26)-(28). Hence, the MSD recurrence of the FC-DNLMS algorithm is given by

$$\tilde{K}_{ii}(n+1) = \tilde{K}_{ii}(n) - \tilde{\alpha}_i(n)\tilde{K}_{ii}(n) + \sum_{r=1, r\neq i}^{N} \tilde{\beta}_{ir}(n)\tilde{K}_{rr}(n) + \tilde{\gamma}_i(n) \tag{65}$$

where

$$\tilde{\alpha}_i(n) = 2\sum_{j=1}^{M} c_j \frac{\xi_j}{N\overline{\sigma_{xj}^2(n)}} \sigma_{xj}^2(n-i+1) - \sum_{j=1}^{M} c_j^2 \left[\frac{\xi_j}{N\overline{\sigma_{xj}^2(n)}}\right]^2 \psi_j \sigma_{xj}^4(n-i+1)$$

$$- \sum_{j=1}^{M}\sum_{k\neq j}^{M} c_j c_k \left[\frac{\xi_j}{N\overline{\sigma_{xj}^2(n)}}\right]\left[\frac{\xi_k}{N\overline{\sigma_{xk}^2(n)}}\right] \sigma_{xj}^2(n-i+1)\sigma_{xk}^2(n-i+1) \tag{66}$$

$$\tilde{\beta}_{ir}(n) = \sum_{j=1}^{M} c_j^2 \left[\frac{\xi_j}{N\overline{\sigma_{xj}^2(n)}}\right]^2 \sigma_{xj}^2(n-i+1)\sigma_{xj}^2(n-r+1) \tag{67}$$

$$\tilde{\gamma}_i(n) = \sum_{j=1}^{M} c_j^2 \left[ \frac{\xi_j}{N\sigma_{xj}^2(n)} \right]^2 \sigma_{nj}^2 \sigma_{xj}^2(n-i+1) + \sigma_q^2. \tag{68}$$

The MSD is given by $MSD(n) = \sum_{i=1}^{N} \tilde{K}_{ii}(n)$.

### 5.3. Behavior for Slowly-Varying Node Input Powers

The case of slowly-varying node input powers is described by (33). In this case, $\overline{\sigma_{xj}^2(n)} \approx \sigma_{xj}^2(n)$.

### 5.3.1 Mean weight behavior

The mean weight behavior of the FC-DNLMS algorithm is obtained from (35) after inserting the substitution (63) with $\overline{\sigma_{xj}^2(n)} \approx \sigma_{xj}^2(n)$. This yields

$$E\{P(n+1)\} = \left[ 1 - \sum_{j=1}^{M} \frac{1}{N} c_j \xi_j \right] E\{P(n)\}. \tag{69}$$

From (69), the transient mean weight behavior is not affected neither by the input distribution type nor by the time variations of the node input powers. The latter is a difference with respect to the FC-DLMS algorithm. If $\xi_j$ is the same for all $j$, denote it by $\xi$, (69) and (4) yield

$$E\{P(n+1)\} = \left[ 1 - \frac{\xi}{N} \right] E\{P(n)\}. \tag{70}$$

Equation (70) has a very interesting implication. For equal step-sizes of the nodes, the mean weight behavior does not depend on the weighting of the nodes, and it is the same as the mean weight behavior of an isolated NLMS algorithm. Eq. (70) implies that the stability range of the mean weight behavior is $0 < \xi < 2N$.

### 5.3.2 MSD behavior

Now, we study the MSD behavior. Inserting the substitution (63), with $\overline{\sigma_{xj}^2(n)} \approx \sigma_{xj}^2(n)$, in (42), the MSD behavior of the FC-DNLMS algorithm is given by

$$MSD(n+1) = \left\{1 - 2\sum_{j=1}^{M}\frac{1}{N}c_j\xi_j + \sum_{j=1}^{M}\frac{1}{N^2}c_j^2\xi_j^2(\psi_j + N - 2)\right.$$

$$\left. + \left[\sum_{j=1}^{M}\frac{1}{N}c_j\xi_j\right]^2\right\}MSD(n) + \sum_{j=1}^{M}\frac{c_j^2\xi_j^2\sigma_{nj}^2}{N\sigma_{xj}^2(n)} + N\sigma_q^2. \quad (71)$$

From (71), the MSD dependence on the kurtosis decreases as $N$ increases. The term multiplying $MSD(n)$ on the right hand side of (71), denote it by $A$, determines the transient MSD behavior of the FC-DNLMS algorithm. This term shows that the transient MSD behavior of the FC-DNLMS algorithm is not affected by the cyclostationarity of the input for slowly varying input powers, which is a difference from the FC-DLMS algorithm. The term $A$ shows that both the convergence speed and stability of the FC-DNLMS algorithm depend on the node kurtosis and weighting. If $\xi_j$ is the same for all $j$, denote it by $\xi$, the term $A$ will be equal to

$$A = 1 - \frac{2\xi}{N} + \frac{\xi^2}{N^2}\sum_{j=1}^{M}c_j^2(\psi_j + N - 2) + \frac{\xi^2}{N^2}. \quad (72)$$

From (72), the convergence speed is maximum when $\sum_{j=1}^{M}c_j^2(\psi_j + N - 2)$ is minimum. From the appendix, with $\eta_j = (\psi_j + N - 2)^{-1}$, this minimum will take place for $c_j$ satisfying

$$c_j = \frac{(\psi_j + N - 2)^{-1}}{\sum_{k=1}^{M}(\psi_k + N - 2)^{-1}} \quad (73)$$

and the minimum value of $\sum_{j=1}^{M}c_j^2(\psi_j + N - 2)$ is given by

$$\left[\sum_{j=1}^{M}c_j^2(\psi_j + N - 2)\right]_{min} = \frac{1}{\sum_{k=1}^{M}(\psi_k + N - 2)^{-1}}. \quad (74)$$

From (72), the step-size stability bound $\xi_o$ is given by

$$\xi_o = \frac{2N}{1+\sum_{j=1}^{M} c_j^2 (\psi_j + N - 2)}. \tag{75}$$

The maximum value of $\xi_o$ is attained at $c_j$ satisfying (73). From (74), the maximum value of $\xi_o$ is given by

$$\xi_{o\max} = \frac{2N \sum_{k=1}^{M}(\psi_k + N - 2)^{-1}}{\sum_{k=1}^{M}(\psi_k + N - 2)^{-1} + 1}. \tag{76}$$

For the same kurtosis for all nodes, the maximum step-size stability bound is equal to

$$\xi_{o\max} = \frac{2NM}{M + \psi + N - 2}. \tag{77}$$

and it is attained at $c_j = 1/M$ for all $j$. Eq. (77) has a very interesting implication. For a single node, $M=1$ and (77) yields

$$\xi_{o\max} = \frac{2N}{\psi + N - 1}. \tag{78}$$

Thus, the stability range for FC is larger than for a single node. Hence the convergence rate for the FC can be faster than for a single node without causing instability. In essence, each of the nodes is helping to stabilize the other nodes. The effects of this property on the MSD will be investigated in Section 6 when comparing the theory and Monte Carlo simulations.

Now, let us investigate the steady-state behavior of the algorithm. Replacing both $MSD(n)$ and $MSD(n+1)$ in (71) by $MSD_{steady-state}(n)$, we obtain

$$MSD_{steady-state}(n) = \left\{ 2\sum_{j=1}^{M} \frac{1}{N} c_j \xi_j - \sum_{j=1}^{M} \frac{1}{N^2} c_j^2 \xi_j^2 (\psi_j + N - 2) - \left[\sum_{j=1}^{M} \frac{1}{N} c_j \xi_j\right]^2 \right\}^{-1}$$

$$\times \left\{ \sum_{j=1}^{M} \frac{c_j^2 \xi_j^2 \sigma_{nj}^2}{N\sigma_{xj}^2(n)} + N\sigma_q^2 \right\}. \tag{79}$$

Eq. (79) implies that slow input power variations generally cause ripples in the steady-state MSD. The ripple repetition period is equal to the least common multiple of the power repetition periods at all nodes.

For wide-sense stationary node inputs, if the step-size is small and the same for all nodes, eq. (79) implies that

$$MSD_{steady-state} = \frac{\xi}{2} \sum_{j=1}^{M} \frac{c_j^2}{\rho_j} + \frac{N^2 \sigma_q^2}{2\xi} \qquad (80)$$

where $\rho_j \equiv \sigma_{xj}^2 / \sigma_{nj}^2$ is the SNR at node $j$. Using the appendix, with $\eta_j = \rho_j$, eq. (80) implies that the minimum steady-state MSD is attained at

$$c_j = \frac{\rho_j}{\rho_1 + \rho_2 + ... + \rho_M} \qquad (81)$$

and the minimum steady-state MSD is given by

$$MSD(\infty)_{min} = \frac{1}{2}\left[\frac{\xi}{\rho_1 + \rho_2 + ... + \rho_M} + N^2 \xi^{-1} \sigma_q^2\right]. \qquad (82)$$

Eq. (82) implies that the steady-state MSD is decreasing as the number of nodes increases and the node signal-to-noise ratios increase. Both behaviors agree with physical intuition.

## 6 MONTE CARLO SIMULATIONS AND COMPARISON TO THEORY

This section presents Monte Carlo simulations of the FC-DLMS and FC-DNLMS algorithms. In all considered cases, the number of nodes is $M=10$, the number of filter taps is $N=32$, $H(0)$ is a two sided decaying exponential with decay factor 0.5. The number of MC simulations is 100. The mean-square increment of the unknown system is $\sigma_q^2 = 64 \times 10^{-8}/N$. The nodal noise power $\sigma_{nj}^2 = 10^{-6}$ for all $j$. The step sizes $\mu_j$ and $\xi_j$ are equal to $\mu$ and $\xi$, respectively, for all nodes, for all figures except Figs. 9 and 10. The weights $c_j$ are equal to $1/M$ for all nodes. Only sinusoidal variations (9) of the nodal input powers are considered. The pulsed case in (10) will not be presented here for reasons of space. Any diffusion results based on the pulsed case are expected to be similar to those for the sinusoidal case. The time-averaged input signal power is unity for all nodes; i.e. $\beta_j = 1$ for all $j$, for all figures except Figs. 9 and 10. To check the behaviors of the algorithms for all rates of variation of the nodal input powers, we used $\omega_{oj} = 2\pi/(2^j)$, $j$ = 1, 2, …, $M$. This corresponds to a rich mixture of fast, intermediate and slow variations.

The three types of non-Gaussian distributions studied have small, medium and large kurtosis. The $j$'th node input is given by (7) with $s_j(n)$ a white uniformly distributed random sequence with zero mean and unity variance with kurtosis 9/5 for the first type. $s_j(n)$ is a white Laplacian sequence with zero mean and unity variance with kurtosis 6 for the second type. For the third type, $s_j(n) = u_j^5(n)/\sqrt{945}$ with $u_j(n)$ a white Gaussian sequence with zero-mean and unity variance. $s_j(n)$ has zero-mean and unity variance with kurtosis 733 for the third type. $s_j(n)$ and $s_k(n)$ are mutually independent for $j \neq k$ in all cases.

Figure 3 shows the results of the FC-DLMS algorithm for uniformly distributed nodal inputs. The figure has two parts (a) and (b) for μ respectively equal to ½ and 2 of the step-size stability bound $2/(N+\psi-1)$ = 0.061 for an isolated node. The purpose of this choice is to support the stability theory of the FC-DLMS algorithm for step-sizes exceeding the stability bound for an isolated node (derived in Section 4). Both parts of Figure 3 show excellent agreement between theory and simulations. This supports the derived theory for all rates of variation of the nodal input powers.

Figure 4 shows the MSD results for the FC-DLMS algorithm for Laplacian distributed nodal inputs for μ = $1/(N+\psi-1)$ = 0.027. Excellent agreement between theory and simulations is obtained.

Figure 5 shows the MSD results for the FC-DLMS algorithm for the Gaussian to the fifth power input with kurtosis = 733 for μ = $1/(N+\psi-1)$ = 0.0013. The agreement between theory and simulations for this exaggerated kurtosis value suggests that the theoretical model holds for all types of distributions for the nodal inputs.

The cases in Figs. 4 and 5 have also been run for μ = $4/(N+\psi-1)$ for each case. The results, not shown here to save space, support the proved stability of the FC-DLMS algorithm for step-sizes exceeding the stability bound for an isolated node.

Figure 6 shows the results for the FC-DNLMS algorithm for uniformly distributed nodal inputs. The figure has two parts (a) and (b) for $\xi$ respectively equal to ½ and 2 of the step-size stability bound $2N/(N+\psi-1) = 1.9512$ for an isolated node (see eq (78)). The purpose of this choice is to support the stability of the FC-DNLMS algorithm for step-sizes exceeding the stability bound for an isolated node (see Section 5). Figure 6 shows excellent agreement between theory and simulations and supports the derived FC-DNLMS theory for all rates of variation of the nodal input powers.

Figure 7 shows the MSD results for the FC-DNLMS algorithm for Laplacian distributed nodal inputs for $\xi = N/(N+\psi-1) = 0.864$. Excellent agreement between theory and simulations is obtained.

Figure 8 shows the results of the FC-DNLMS algorithm for the fifth power of Gaussian input whose kurtosis is equal to 733 for $\xi = N/(N+\psi-1) = 0.0419$. Note that even in this case where condition (61) is clearly violated, there is only a small mismatch between theory and simulations for this exaggerated value of the kurtosis. This suggests that the derived theoretical model holds for all types of distributions of the nodal inputs, with sufficient accuracy for design purpose even in extreme cases.

The cases in Figs. 7 and 8 have also been run for $\xi = 4N/(N+\psi-1)$ for each case. The results, not shown here to save space, support the proved stability of the FC-DNLMS algorithm for step-sizes exceeding the stability bound for an isolated node.

Finally, Figs. 9 and 10 show the results for the FC-DLMS and FC-DNLMS algorithms respectively for a case in which input powers, input kurtosis and step sizes are different for different nodes. The input is uniformly distributed for nodes 1-4, Laplacian for nodes 5-7, and fifth power of Gaussian for nodes 8-10. The time-averaged input power for the *j*-th node is $\beta_j = 1$ for $1 \le j \le 5$ and $\beta_j = 0.1$ for $6 \le j \le 10$. The *j*-th step size is $\mu_j = 1/[\beta_j(N + \psi_j - 1)]$ for the DLMS algorithm and $\xi_j = N/(N + \psi_j - 1)$ for the DNLMS algorithm. Thus, the step-size vectors for the DLMS and DNLMS algorithms are equal to [0.0305, 0.0305, 0.0305, 0.0305, 0.0270, 0.270, 0.270, 0.013, 0.013, 0.013] and [0.976, 0.976, 0.976, 0.976, 0.864, 0.864, 0.864, 0.0416, 0.0416, 0.0416] respectively. As mentioned at the beginning of this section, the rates of variation of the nodal input powers are different for different nodes. The agreement between theory and simulations in Figs. 9 and 10 suggest that the derived model is applicable for all scenarios of the network.

**7 CONCLUSIONS**

This paper has derived a relatively simple theory for the DLMS and DNLMS algorithms for a network having a fusion center. The analysis is done in a system identification framework for cyclostationary white nodal inputs. The analysis holds for all types of nodal input distributions and nodal input power variations. The derived models consist of simple scalar recursions. These recursions facilitate the derivation of interesting conclusions about the behaviors of the algorithms: 1) The MSD behavior depends on the node input distribution through only the kurtosis for both the DLMS and DNLMS algorithms. This implies that different input distributions with the same kurtosis yield the same MSD. 2) The MSD dependence on the kurtosis increases as the step-size increases and as the filter length $N$ decreases. 3) The mean weight behavior of the DLMS (DNLMS) algorithm depends (does not depend) on the time variations of the nodal input powers and weighting coefficients of the nodes. 4) Slow input power variations cause (do not cause) ripples in the transient MSD of the DLMS (DNLMS) algorithm. 5) The convergence rate is a decreasing function of both the kurtosis and the sum of the squared node weighting coefficients for both the DLMS and DNLMS algorithms. Fastest convergence is attained for uniform weighting coefficients. 6) The MSD dependence on the kurtosis decreases as the number of nodes $M$ increases. The MSD stability depends on the nodal kurtosis and nodal weighting coefficients. The maximum step-size stability bound is attained at equal nodal weighting coefficients and increases as $M$ increases. The bound exceeds the step-size stability bound for an isolated node. 7) Slow input power variations generally cause ripples in the steady-state MSD. The ripple repetition period is equal to the least common multiple of the power repetition periods at all nodes. The ripples weaken as $M$ increases. 8) The steady-state MSD decreases as $M$ increases and the node signal-to-noise ratios increase. Analytical results are in excellent agreement with simulations.

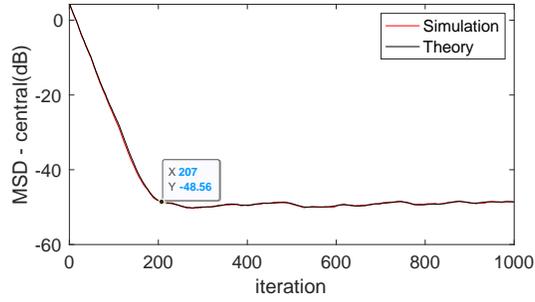

(a)

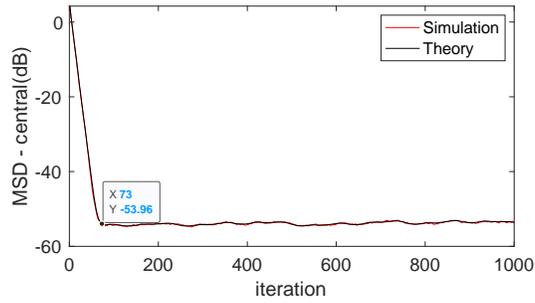

(b)

Figure 3: FC-DLMS, uniformly distributed inputs, (a) $\mu=1/(N+\psi-1) = 0.0305$, (b) $\mu=4/(N+\psi-1) = 0.122$.

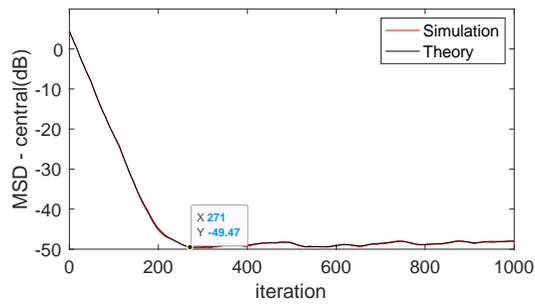

Figure 4: FC-DLMS, Laplacian inputs, $\mu=1/(N+\psi-1) =0.0270$.

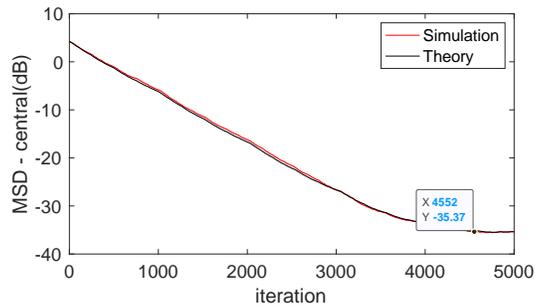

Figure 5: FC-DLMS, Gaussian to the fifth power inputs, $\mu=1/(N+\psi-1) = 0.0013$.

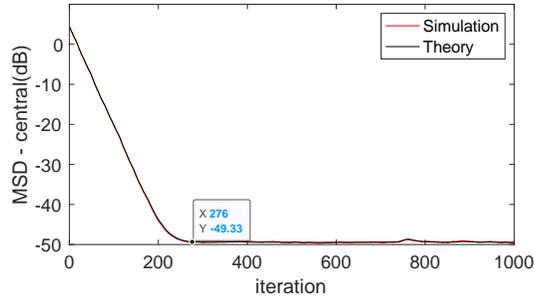

(a)

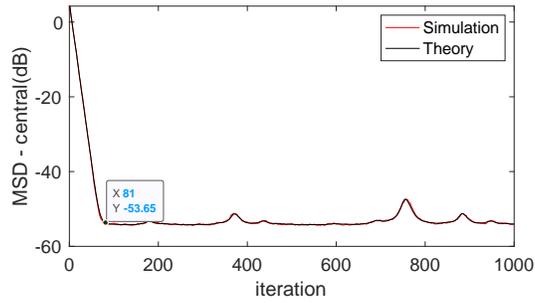

(b)

Figure 6: FC-DNLMS, uniform inputs, (a) $\xi = N/(N+\psi-1) = 0.9756$, (b) $\xi = 4N/(N+\psi-1) = 3.9024$.

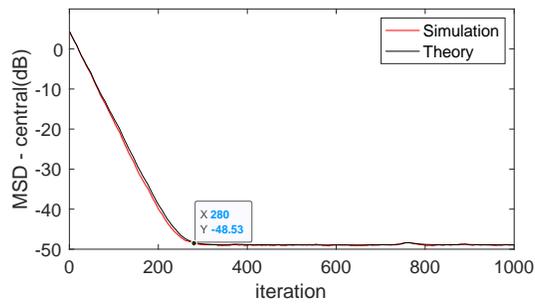

Figure 7: FC-DNLMS, Laplacian inputs, $\xi = N/(N+\psi-1) = 0.8649$.

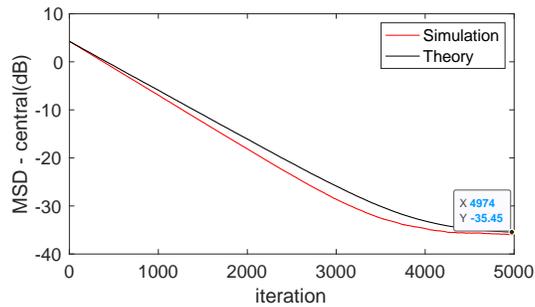

Figure 8: FC-DNLMS, Gaussian to the fifth power inputs, $\xi = N/(N+\psi-1) = 0.0419$.

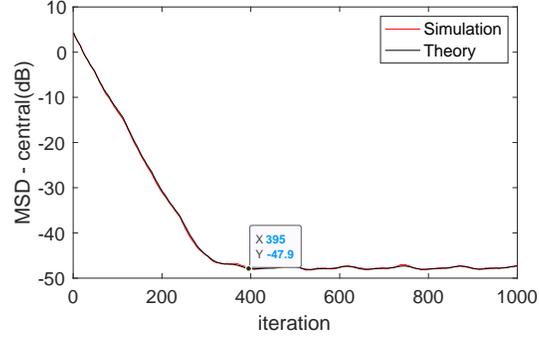

Figure 9: FC-DLMS, different node input powers, input kurtosis and step sizes. Step-size vector = [0.0305, 0.0305, 0.0305, 0.0305, 0.0270, 0.270, 0.270, 0.013, 0.013, 0.013].

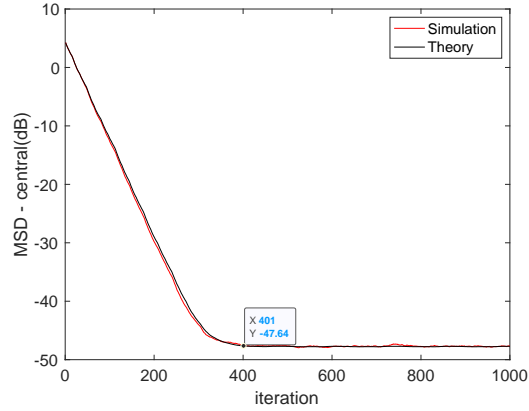

Figure 10: FC-DNLMS, different node input powers, input kurtosis and step sizes. Step-size vector = [0.976, 0.976, 0.976, 0.976, 0.864, 0.864, 0.864, 0.0416, 0.0416, 0.0416].

## APPENDIX

$$\text{Minimum of } \sum_{j=1}^{M} \frac{c_j^2}{\eta_j}$$

For $\eta_j > 0,\ j = 1, 2, ..., M$, denote

$$f \equiv \sum_{j=1}^{M} \frac{c_j^2}{\eta_j}. \tag{83}$$

From (4), we have

$$f = \sum_{j=1}^{M-1} \frac{c_j^2}{\eta_j} + \frac{1}{\eta_M} \left[ 1 - \sum_{j=1}^{M-1} c_j \right]^2. \tag{84}$$

Differentiating eq. (84) on $c_k$ for $k=1, 2.,..M$ yields

$$\frac{\partial f}{\partial c_k} = \frac{2c_k}{\eta_k} - \frac{2}{\eta_M} \left[ 1 - \sum_{j=1}^{M-1} c_j \right] = 0 \tag{85}$$

for a minimum or maximum of $f$ and again to determine if the point is a maximum or minimum

$$\frac{\partial^2 f}{\partial c_k^2} = \frac{2}{\eta_k} + \frac{2}{\eta_M} > 0. \tag{86}$$

From (85) and (86), $f$ is minimum for

$$\frac{c_k}{\eta_k} = \frac{1}{\eta_M} \left[ 1 - \sum_{j=1}^{M-1} c_j \right] = \frac{c_M}{\eta_M}, \quad k = 1, 2, ..., M-1. \tag{87}$$

Eq. (87) implies that $f$ is minimum for

$$c_k = \frac{\eta_k}{\sum_{j=1}^{M} \eta_j} \tag{88}$$

and that the minimum value of $f$ is equal to

$$f_{\min} = \frac{1}{\sum_{j=1}^{M} \eta_j}. \tag{89}$$